\documentclass[prd,preprint,superscriptaddress,showpacs,nofootinbib,%
tightenlines ]{revtex4}
\usepackage{epsfig}
\usepackage{amssymb}

\newcommand{\ben}{\begin{displaymath}}
\newcommand{\een}{\end{displaymath}}
\newcommand{\be}{\begin{equation}}
\newcommand{\ee}{\end{equation}}
\newcommand{\bea}{\begin{eqnarray}}
\newcommand{\eea}{\end{eqnarray}}
\begin{document}
%\preprint{MKPH-T-06-16}
\title{The magnetic moment of the {\boldmath$\rho$}-meson}
\author{D.~Djukanovic}
\affiliation{Helmholtz Institute Mainz, University of Mainz, D-55099 Mainz, Germany}
\author{E.~Epelbaum}
 \affiliation{Institut f\"ur Theoretische Physik II, Ruhr-Universit\"at Bochum,  D-44780 Bochum,
 Germany}
\author{J.~Gegelia}
 \affiliation{Institut f\"ur Theoretische Physik II, Ruhr-Universit\"at Bochum,  D-44780 Bochum,
 Germany}  \affiliation{Tbilisi State  University,  0186 Tbilisi,
 Georgia}
\author{U.-G.~Mei\ss ner}
 \affiliation{Helmholtz Institut f\"ur Strahlen- und Kernphysik and Bethe
   Center for Theoretical Physics, Universit\"at Bonn, D-53115 Bonn, Germany}
 \affiliation{Institute for Advanced Simulation, Institut f\"ur Kernphysik
   and J\"ulich Center for Hadron Physics, Forschungszentrum J\"ulich, D-52425 J\"ulich,
Germany}
 \date{16 September 2013}
\begin{abstract}
The magnetic moment of the $\rho$-meson is calculated in the framework
of a low-energy effective field theory of the strong interactions. We
find that the complex-valued strong interaction  corrections to the gyromagnetic
ratio are small leading to a value close to the real leading tree level
result, $g_\rho  = 2$. This is in a reasonably good agreement with the available
lattice QCD calculations for this quantity.

%{\bf The results given in this draft are correct, crosschecked by
%Dalibor and Jambul}

\end{abstract}

% insert suggested PACS numbers in braces on next line

%\pacs{
%03.65.Nk,
%Nonrelativistic scattering theory
%11.10.Gh,
%Renormalization
%12.39.Fe.}
%Chiral Lagrangians
%04.60.Ds  Canonical quantization
%04.60.Gw  Covariant and sum-over-histories quantization
%03.70.+k  Theory of quantized fields
%          (see also 11.10 Field theory)


%\pacs{

\maketitle

\section{introduction}

   Phenomenological low-energy chiral Lagrangians with vector mesons
were considered already in the 1960s, see e.g.
Refs.~\cite{Schwinger:1967tc,Wess:1967jq,Weinberg:de}.
Later, the chiral symmetry of QCD has been taken into account
in the framework of effective field theories describing the
interaction of vector mesons with
pseudoscalars and baryons, see, e.g.,
Refs.~\cite{Gasser:1984yg,Meissner:1987ge,Bando:1988br,
Ecker:yg,Ecker:1988te,Borasoy:1996ds,Birse:1996hd,Kubis:2000zd,Harada:2003jx,Bruns:2004tj,Bruns:2008ub}.

In a covariant formalism, massive vector bosons are described by
Lagrangians with constraints. The self-consistency of a system
with constraints imposes non-trivial conditions on the form of the Lagrangian.
In Ref.~\cite{Djukanovic:2004mm}, the effective Lagrangian of
Ref.~\cite{Weinberg:de} describing the interaction among
$\rho$-mesons, pions and nucleons was considered. Requiring
perturbative renormalizability in the sense of effective field
theory \cite{Weinberg:mt},  the universality of the vector-meson
couplings was derived. The crucial ingredient of any effective field
theory (EFT) is power counting. It is possible to consistently
include virtual (axial-) vector mesons
in EFT \cite{Kubis:2000zd,Fuchs:2003sh,Schindler:2003xv} provided they appear
only as internal lines in Feynman diagrams involving soft external
pions and nucleons with small three-momenta. The issue becomes
highly non-trivial for energies when the intermediate resonant
states can be generated. The problem is that vector mesons decay in light
modes and therefore large imaginary parts appear \cite{Bruns:2004tj}.
First attempts have been made to handle this problem
by applying the complex mass scheme 
\cite{Stuart:1990,Denner:1999gp,Djukanovic:2009zn,Djukanovic:2009gt,Djukanovic:2010id,Bauer:2011bv,Bauer:2012at,Bauer:2012gn}.

    In this work we calculate the magnetic moment of the $\rho$-meson
as a function of quark masses in the framework of a low-energy effective
theory of the strong interactions. As for any unstable particle, this
quantity is a complex number. For a detailed discussion on this issue,
 see Ref.~\cite{Gegelia:2009py}.
    We start with the most general effective Lagrangian of vector mesons
interacting with pions in the presence of external fields.
   It contains an infinite number of interaction terms which
respect  the underlying symmetries of QCD.
In this work, we make the assumption that the interaction terms with
a higher number of derivatives and/or
more fields are suppressed by powers of some large hadronic
scale. Therefore, only a finite number of terms of the effective
Lagrangian is required to achieve a given accuracy.
We apply the complex mass renormalization scheme
\cite{Stuart:1990,Denner:1999gp} and
calculate the magnetic moment of the vector meson  and its pion mass
dependence at one-loop order.

\section{Lagrangian}

We start with the most general chiral effective Lagrangian for $\rho$- and
$\omega$-mesons, pions and external sources in the parametrization of the model~III
of Ref.~\cite{Ecker:yg}\footnote{The $\rho$-vector fields
transform inhomogeneously under chiral transformations in this
parametrization.
This coincides with the parametrization adopted by Weinberg for the
vector fields in Ref.~\cite{Weinberg:de}. }
\begin{displaymath}
{\cal L}={\cal L}_{\pi}+{\cal L}_{\rho\pi}+{\cal L}_\omega
+{\cal L}_{\omega\rho\pi}+\cdots~.
\end{displaymath}
Below we specify only those terms of the Lagrangian which are relevant
for the calculation of the magnetic moment of the $\rho$-meson presented in this work:
\begin{eqnarray}
{\cal L}_\pi& = & \frac{F^2}{4}\,{\rm Tr} \left[D_\mu U
\left(D^\mu
U\right)^\dagger\right]+\frac{F^2\,M^2}{4}\,{\rm Tr} \left(
U^\dagger+U\right), \nonumber\\
{\cal L}_{\rho\pi}&=&
- \frac{1}{2}\,{\rm
Tr}\left(\rho_{\mu\nu}\rho^{\mu\nu}\right) + \left[ M_{\rho}^2 +
\frac{c_{x}\,M^2\,{\rm Tr} \left(
U^\dagger+U\right) }{4}\right]
{\rm Tr}\left[\left(
\rho^\mu-\frac{i\,\Gamma^\mu}{g}\right)\left(
\rho_{\mu}-\frac{i\,\Gamma_\mu}{g} \right)\right],\nonumber\\
%&+& i\,d_x {\rm
%Tr}\left[\rho_{\mu\nu}\Gamma^{\mu\nu}\right]
%%-\frac{1}{2}\,{\rm
%%Tr}\left[{\cal V}_{\mu\nu}{\cal
%%V}^{\mu\nu}\right]
%%\nonumber\\
%- \frac{\sqrt{2}}{2}f_V{\rm
%Tr}\left\{\rho_{\mu\nu}f_+^{\mu\nu}\right\},\nonumber\\
{\cal L}_\omega&=&
 -\frac{1}{4}\left(
\partial_\mu\omega_{\nu}-\partial_\nu\omega_{\mu}\right)\left(
\partial^\mu\omega^\nu-\partial^\nu\omega^\mu\right)+\frac{M_{\omega}^2\,
\omega_{\mu}\omega^\mu}{2},\nonumber\\
%&& +
%g_{\omega\rho\pi}\epsilon_{\mu\nu\alpha\beta}\left(\partial^\mu
%\omega^{\nu}\right) \left(\partial^\alpha \rho^{a \beta}\right)
%\pi^a \nonumber\\
{\cal L}_{\omega\rho\pi}&=&\frac{1}{2}\,
g_{\omega\rho\pi}\,\epsilon_{\mu\nu\alpha\beta}\, \omega^{\nu}\,
{\rm Tr}\left(\rho^{\alpha\beta} u^\mu
\right),\label{finallagrangian}
\end{eqnarray}
where
\begin{eqnarray}
U&=&u^2={\rm exp}\left(\frac{i\vec{\tau}\cdot\vec{\pi}}{F}\right),
\nonumber\\
\rho^\mu & = & \frac{\vec\tau\cdot\vec\rho\,^\mu}{2},\nonumber\\
%\rho_0^{a,\mu} \,\frac{\tau^a}{2}\,,\nonumber\\
\rho^{\mu\nu} & = &
\partial^\mu\rho^\nu-\partial^\nu\rho^\mu - i
g\left[\rho^\mu,\rho^\nu\right] \,,\nonumber
\\
%\rho_0^{\mu\nu} & = &
%\partial^\mu\rho_0^\nu-\partial^\nu\rho_0^\mu - i
%g\left[\rho_0^\mu,\rho_0^\nu\right] \,,\nonumber\\
u_\mu & = & i \left[ u^\dagger \partial_\mu u-u \partial_\mu
u^\dagger -i(u^\dagger v_\mu u-u v_\mu
u^\dagger)\right], \nonumber\\
\Gamma_\mu &= & \frac{1}{2}\,\biggl[ u^\dagger\partial_\mu u+u
\partial_\mu u^\dagger
- i\,\left( u^\dagger v_\mu u +u v_\mu
u^\dagger\right)\biggr]~,\nonumber
\\
\Gamma_{\mu\nu} & = &
\partial_\mu \Gamma_\nu-\partial_\nu \Gamma_\mu
+[\Gamma_\mu,\Gamma_\nu]\,, \nonumber
\\
f^{\mu\nu}_{+} & = & u F^{\mu\nu}_L u^\dagger + u^\dagger
F^{\mu\nu}_R u\,, \nonumber
\\
D_\mu A & = & \partial_\mu A -i v_\mu A+i A v_\mu\,.
\label{somedefinitions}
\end{eqnarray}
Here, $F$ denotes the pion decay constant in the chiral
limit, $M^2$ is the lowest order expression for the
squared pion mass, $M_\rho$ and $M_\omega$ refer to the $\rho$ and
$\omega$ masses in the chiral limit, respectively. Further,
$g$, $c_x$
%, $d_x$, $f_V$
and $g_{\omega\rho\pi}$ are coupling constants and $v_\mu$ is the
external vector field.
Notice that we do not show the counterterms explicitly. For the electromagnetic
interaction we have $v_\mu=-e\, \tau^3 A_\mu/2$.
   Demanding that couplings with different mass dimensions are
independent, the consistency condition for the $\rho\pi\pi$
coupling \cite{Djukanovic:2004mm} leads to the KSFR relation
\cite{Kawarabayashi:1966kd,Riazuddin:sw}
\begin{eqnarray}
M_\rho^2 & = & 2\,g^2 F^2 \,.\label{M0}
\end{eqnarray}

\section{Magnetic moment of the vector meson}
\label{sec:mmvm}

\begin{figure}
\epsfig{file=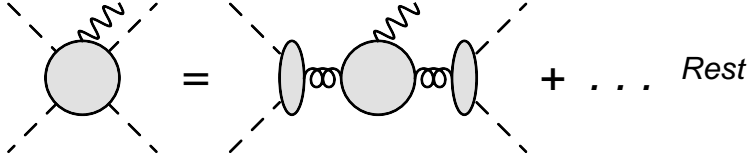,width=0.6\textwidth}
\caption[]{\label{QFTPR} The reaction  $\pi \pi\to
  \gamma\pi\pi$. Dashed, wavy and wiggled lines refer to pions,
  photons and $\rho$-mesons, respectively. Non-resonant contributions
  denoted by ``Rest'' are not shown explicitly.}
\end{figure}

As the $\rho$-meson is an unstable particle it does not appear as an asymptotic state in
the effective field theory. Therefore, to define the magnetic moment of the $\rho$-meson,
we follow the strategy of Ref.~\cite{Gegelia:2009py} and consider an
amplitude of a process in which the $\gamma \rho\rho$
vertex contributes as a sub-diagram. For the sake of definiteness, we take
the process $\pi \pi\to \gamma\pi\pi$ shown in Fig.~\ref{QFTPR}.
We parameterize the amplitude of this process as
\begin{equation}
{\cal M} = {\cal M}_1^\alpha\, (-i) \,D_{\alpha\mu}(p_i)
V^{\lambda\mu\nu}(q,p_f,p_i)\, (-i) \,D_{\nu\beta}(p_f){\cal
M}_2^\beta \epsilon_\lambda + { Rest}\,,
\label{completeprocess}
\end{equation}
where ``{\it Rest}'' denotes the non-resonant contributions.
\noindent
Here, ${\cal M}_1^\alpha$ and ${\cal M}_2^\beta$ are the $\rho\pi\pi$
vertex functions, $\epsilon_\lambda$ is the photon polarization, and $-i D^{\mu\nu} (p)$ with
\begin{eqnarray}
D^{\mu\nu} (p)& = & Z_V \, \frac{g^{\mu\nu}-p^\mu
p^\nu/z}{p^2-z} + R
\label{DPparameterized}
\end{eqnarray}
is the dressed propagator of the vector meson. Further,  $Z_V$  is
the (complex) residue at the pole $z$ and $R$
denotes the non-pole part.
The $\gamma \rho\rho$ vertex function can
be written as
\begin{eqnarray}
V^{\lambda\mu\nu}(q,p_f,p_i) & = & \sum_j t_j^{\lambda\mu\nu}
V_j(q^2,p_f^2,p_i^2) \,,
\label{DVparameterized}
\end{eqnarray}
where $t_j^{\lambda\mu\nu}$ denote the possible tensor
structures which depend on the momenta $p_{i}$, $p_f$  and the metric
tensor, and $V_i (q^2,p_f^2,p_i^2) $ are the corresponding scalar
functions. Here and in what follows, we do not show the isospin
indices (unless stated otherwise) for the sake of compactness. Expanding
the $V_j$
about the $z$-pole and substituting, together with the expression for
the propagator in
Eq.~(\ref{DPparameterized}), into Eq.~(\ref{completeprocess}) we
obtain for the leading double-pole contribution
\begin{equation}
{\cal M}_{\rm dp} = - {\cal
M}_1^\alpha\,\frac{g^{\alpha\mu}-p_i^\mu
p_i^\nu / z}{p_i^2-z}\,Z_V \sum_j t_j^{\lambda\mu\nu}
V_j(q^2,z,z)\,Z_V\, \frac{g^{\nu\beta}-p_f^\nu
p_f^\beta / z}{p_f^2-z}\,{\cal M}_2^\beta\,.
\label{completeprocessleading}
\end{equation}
In order to properly renormalize the $\gamma \rho\rho$ vertex function we rewrite
Eq.~(\ref{completeprocessleading}) in the form
\begin{equation}
{\cal M}_{\rm dp} = - {\cal M}_1^\alpha\,\sqrt{Z_V}
\frac{g^{\alpha\mu}-p_i^\mu
p_i^\nu / z}{p_i^2-z}\,\sqrt{Z_V} \sum_j t_j^{\lambda\mu\nu}
V_j(q^2,z,z)\,\sqrt{Z_V}\, \frac{g^{\nu\beta}- p_f^\nu
p_f^\beta / z}{p_f^2-z}\,\sqrt{Z_V}\,{\cal M}_2^\beta
\label{completeprocessleadingrewritten}
\end{equation}
and define
\begin{equation}
i
\,\Gamma^{\lambda\mu\nu}(q,p_i,p_f) := \sqrt{Z_V} \sum_j t_j^{\lambda\mu\nu}
V_j(q^2,z,z)\,\sqrt{Z_V}\,. \label{vertexfunction}
\end{equation}
Noting that $p^\mu D_{\mu\nu}$ does not have a pole, we drop
structures containing $p_f^\nu$ and $p_i^\mu$ and parameterize the
``on-mass-shell'' $\Gamma$ as follows
\begin{equation}
\Gamma^{\lambda\mu\nu}(q,p_i,p_f) = f_1(q^2) \left(
p_i^\lambda+p_f^\lambda\right) g^{\mu\nu}+f_2(q^2) \left(q^\nu
g^{\lambda \mu}-q^\mu g^{\lambda\nu}\right)+\cdots\,,
\label{gvvvertex}
\end{equation}
where the ellipsis refer to structures which do not
involve the metric tensor.

%\begin{figure}
%\epsfig{file=RhoSE.eps,width=0.85\textwidth}
%\caption[]{\label{twoloopSE:fig} Vector meson self-energy
%diagrams at the one-loop level. Wiggly, dashed and solid lines
%correspond to $\rho$, $\pi$ and $\omega$ mesons, respectively.}
%\end{figure}

\medskip

The charge and  magnetic moment $e$ and $\mu_\rho$ of the $\rho$-meson
are defined in terms of the corresponding form factors $f_1(0)$
and $f_2(0)$ as
\begin{eqnarray}
f_1(0) & = & -e\,, \nonumber\\
f_2(0) & = & -2\,M_\rho \mu_\rho\,. \label{mmdef}
\end{eqnarray}

%\newpage
\begin{figure}
\epsfig{file=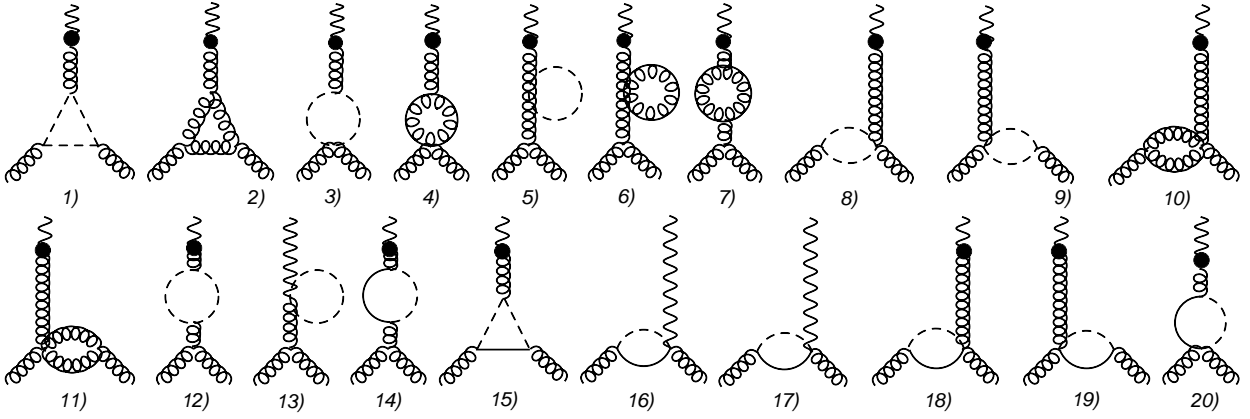,width=1.\textwidth}
\caption[]{\label{OneloopRMM:fig} Leading one loop diagrams contributing
to the magnetic moment of the $\rho$-meson. Wavy, wiggly, dashed and
solid lines correspond to the photon and $\rho$, $\pi$ and $\omega$-mesons,
respectively. The solid circle corresponds to the photon-$\rho$-meson mixing.}
\end{figure}

There are both tree-level and loop contributions to these
quantities. Loop diagrams are suppressed by powers of $\xi = g_i^2/(16
\pi^2)$, where $g_i$ stands for coupling constants in general.
Even for a sizeable coupling like $g_{\rho\pi\pi}$, this expansion
parameter is small, $\xi \simeq 0.2$.
Vertices generated by the $c_x$-term of the Lagrangian are
only included at tree order in our calculations as their contributions
are suppressed at the one-loop order by two additional powers of the pion mass.
At tree order, we obtain
\begin{eqnarray}
f_1^{\rm tree}(0) & = & -e\,,\nonumber\\
f_2^{\rm tree}(0) & = & -2\,e\,,
\label{treecontr}
\end{eqnarray}
in agreement with the findings of Refs.~\cite{Djukanovic:2005ag,Lorce:2009bs}.
One-loop diagrams contributing to the $\gamma\rho\rho$ vertex function
are shown in Fig.~\ref{OneloopRMM:fig}.
Their full contributions to $f_2(0)$ are given in the appendix.
Taking into account the wave function renormalization we obtain
\begin{eqnarray}
f_1^{\rm loop}(0) & = & 0\,,\nonumber\\
f_2^{\rm loop}(0) & = &\frac{e}{384 \pi^2 F^2} \Biggl\{24 M^2
   B_0\left(M_\rho^2,M^2,M^2\right)+30 M_\rho^2
   B_0\left(M_\rho^2,M_\rho^2,M_\rho^2\right)\nonumber\\
&-& 30 A_0\left(M_\rho^2\right)-24 A_0\left(M^2\right)+17 M_\rho^2
   +24 M^2\Biggr\} \nonumber\\
&+& \frac{e \,g_{\omega \rho \pi }^2}{576 \pi ^2 F^2
   M_\rho^2} \Biggl\{12
   A_0\left(M_\omega^2\right) \left(2 M_\omega^2-M_\rho^2-2
   M^2\right)-6 A_0\left(M^2\right)
   \left(4 M_\omega^2+5 M_\rho^2-4 M^2\right)\nonumber\\
   &-& 9
   M_\rho^2 \left(M_\omega^2+M^2\right)+M_\rho^2 \left(21 M_\omega^2-19 M_\rho^2+21
   M^2\right)\nonumber\\
&-& 6 \left[4
   M_\omega^4-2 M^2 \left(4
   M_\omega^2+M_\rho^2\right)+M_\omega^2 M_\rho^2-2
   M_\rho^4+4 M^4\right]
   B_0\left(M_\rho^2,M_\omega^2,M^2\right)\Biggr\}\nonumber\\
&+& \frac{e \,g_{\omega \rho \pi }^2
   \left[M_\omega^2 A_0\left(M_\omega^2\right)-M^2
   A_0\left(M^2\right)\right]}{32 \pi ^2 F^2
   \left(M_\omega^2-M^2\right)}
   \,,
\label{alldiagrams}
\end{eqnarray}
where the loop functions $A_0(m^2)$ and $B_0(p^2,m_1^2,m_2^2)$ are
also defined in the appendix.
Note that in Eq.~(\ref{alldiagrams}), the quantity
$f_1^{\rm loop} (0)$ only vanishes when the complex residue of the
dressed propagator at the pole is used as the wave
function renormalization constant for the vector field.

We now estimate numerically the derived one-loop contributions. Using
$g_{\omega\rho\pi}=1.478$ from Ref.~\cite{Lublinsky:1996yf} and
adopting the physical values for the various meson masses and the pion decay
constant instead of the corresponding chiral-limit values as appropriate at the order we are working, namely
 $M_\rho= 0.775$~${\rm GeV}$,
$M_\omega= 0.782$~${\rm GeV}$, $M \sim M_\pi= 0.1395$~${\rm GeV}$,
$F \sim F_\pi=0.0924$~GeV, we obtain
\begin{equation}
f_2^{loop}(0) = (0.2416-0.0423 \,i) \, e\,. \label{loopnumerical1}
\end{equation}
Notice that using the complex values $M_\rho^2=
(0.775^2-i\,0.775\times 0.149)$~${\rm GeV}^2$, $M_\omega^2=
(0.782^2-i\,0.782\times 0.0085)$~${\rm GeV}^2$ for the renormalized
masses of the $\rho$- and $\omega$-mesons corresponding to the pole
positions leads to a very similar numerical result of
\begin{equation}
f_2^{loop}(0) = (0.2124-0.0415\,i)\, e \,.
\label{loopnumerical2}
\end{equation}

\begin{figure}[t]
\epsfig{file=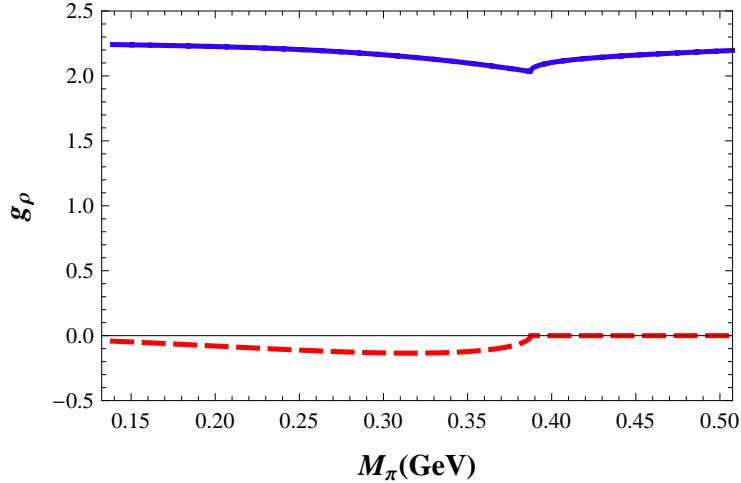,width=0.6\textwidth}
\caption[]{\label{grho} The factor $g_\rho$ as a function of the
pion mass. Solid (blue) and dashed (red) lines correspond to the real
and imaginary parts, respectively.}
\end{figure}

\noindent
Comparing Eqs.~(\ref{loopnumerical1}) and (\ref{loopnumerical2})
with Eq.~(\ref{treecontr}) we see that the loop contributions are
clearly suppressed in comparison with the tree-level result and also
that the imaginary part of the vector meson masses has a little impact
on the value of the loop correction. % by a factor $\sim \Gamma_\rho/M_\rho$.
We thus conclude that the leading quantum corrections to the
classical value of the $g$-factor, $g_\rho = 2$, are suppressed.  This
gives a strong indication that the strong corrections to this
observable are small. This
conjecture is further supported by the available lattice QCD
calculations for this quantity, namely $(g_\rho )_{\rm quenched}\sim 2.3$
of Ref.~\cite{Hedditch:2007ex}  and $(g_\rho )_{\rm
  unquenched}=1.6(1)$ of Ref.~\cite{Gurtler:2008zz} (see also
Ref.~\cite{Andersen:1996qb} for an early study).

Finally, we also
plot in Fig.~\ref{grho}  the real and imaginary parts of the g-factor
$g_\rho$ as functions of the pion mass. Both real and imaginary
part show very little pion mass dependence,
the cusp appears at the value of the pion mass, at which the $\rho$
pole moves from
the second to the first Riemann sheet.
%the cusp is due to the fact
%that we neglected the pion mass dependence of the $\rho$-meson mass.
%However, we do not expect a significant effect from the inclusion of this
%effect. In any case, at these pion masses,  higher order effects should
%also be included.

\section{Summary}

In this paper, we have calculated the complex-valued magnetic moment
of the $\rho$-meson in a chiral effective field theory utilizing the
complex-mass renormalization scheme. Assuming that the interaction
terms with a higher number of derivatives and/or
more fields are suppressed by powers of some large hadronic
scale, we perform a one-loop calculation in terms of the
expansion parameter $\xi = (g_{\rho\pi\pi}/4\pi)^2 \simeq 0.2$. The pertinent
results of our investigation can be summarized as follows:
\begin{itemize}
\item At tree level (leading order), the magnetic moment of the $\rho$ is
  real and its gyromagnetic ratio is $g_\rho = 2$.
\item At one-loop order, the magnetic moment picks up an imaginary part.
We find that the one-loop corrections to  $g_\rho$ are of the order of
$10\%$, cf. Eqs.~(\ref{loopnumerical1},\ref{loopnumerical2}), and the imaginary
part is about 0.04 (in units of the charge). The results are in agreement
with recent lattice QCD determinations.
\item We find that the pion mass dependence of the gyromagnetic ratio
is very weak. This could be tested on the lattice for sufficiently small
pion masses, say $M_\pi \lesssim 0.3\,$GeV, that also allow for the $\rho$-meson
to decay.
\end{itemize}

\acknowledgments

This work was supported in part by Georgian Shota Rustaveli National
Science Foundation (grant 11/31),
DFG (SFB/TR 16,
``Subnuclear Structure of Matter''), by the European
Community-Research Infrastructure Integrating Activity ``Study of
Strongly Interacting Matter'' (acronym HadronPhysics3,
Grant Agreement n. 283286) under the Seventh Framework Programme of EU,
and ERC project 259218 NUCLEAREFT.

\begin{appendix}
\section{Loop functions and explicit expressions}

The loop functions $A_0$ and $B_0$ are defined as follows:
\begin{eqnarray}
A_0(m^2) & = & \frac{(2 \pi)^{4-n}}{i\,\pi^2}\,\int \frac{
d^nk}{k^2-m^2}\,, \nonumber\\
B_0(p^2,m_1^2,m_2^2) & = & \frac{(2 \pi)^{4-n}}{i\,\pi^2}\,\int
\frac{d^nk}{\left[k^2-m_1^2\right]\left[(p+k)^2-m_2^2\right]}\,,
\label{oneandtwoPF}
\end{eqnarray}
where $n$ is the space-time dimension.

\medskip

The sum of all one-particle-irreducible diagrams contributing in
the vector meson two-point function can be parameterized as
\begin{equation}
i\,\Pi^{ab}_{\mu\nu}(p)=i\,\delta^{ab}\left[
g_{\mu\nu}\Pi_{1}+\left(g_{\mu\nu} p^2-p_\mu
p_\nu\right)\,\Pi_2(p^2)\right], \label{VSEpar}
\end{equation}
where $\Pi_1$ is momentum-independent and $\Pi_2(p^2)$ is regular
at $p^2=0$. Further, $a$ and $b$ are the isospin indices.  The wave function
renormalization of the vector meson is
defined as the (complex) residue at the (complex) pole of the
dressed propagator. In terms of Eq.~(\ref{VSEpar}) it reads
\begin{equation}
Z_V  =  \frac{1}{1-\Pi_2(z)-z \Pi_2'(z)}=1+\delta Z_\rho+\cdots\,, \label{residue}
\end{equation}
where $\delta Z_\rho$ is the LO one-loop contribution and  the ellipses
stand for higher order corrections.

\begin{figure}
\epsfig{file=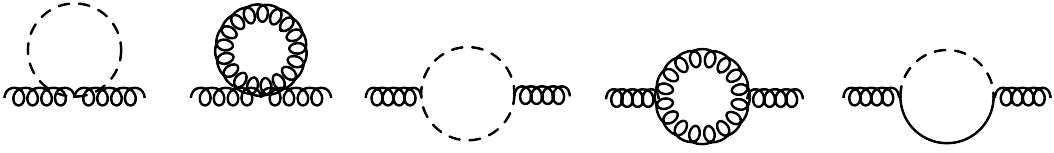,width=0.85\textwidth}
\caption[]{\label{twoloopSE:fig} Vector meson self-energy
diagrams at the one-loop level. Wiggly, dashed and solid lines
correspond to $\rho$, $\pi$ and $\omega$ mesons, respectively.}
\end{figure}

Calculating the vector-meson self-energy diagrams, cf. Fig.~\ref{twoloopSE:fig},
we obtain
%(By Dalibor file ''SigmaPrime'' in Directory ''RhoMagneticMoment'')
\begin{eqnarray}
\delta Z_\rho & = & -\frac{1}{1152 \pi ^2 F^2
   }\Biggl\{4
   \left[3 \left(M_\rho^2+2 M^2\right)
   B_0\left(M_\rho^2,M^2,M^2\right)-6
   A_0\left(M^2\right)-M_\rho^2+6 M^2\right]\nonumber\\
&+& M_\rho^2 \left[99
   B_0\left(M_\rho^2,M_\rho^2,M_\rho^2\right)+113\right]-
   258 A_0\left(M_\rho^2\right)\Biggr\}\nonumber\\
 & + &  \frac{g_{\omega \rho \pi}^2}{288 \pi ^2 F^2
   M_\rho^2} \Biggl\{3 M_\omega^2 M_\rho^2
   B_0\left(M_\rho^2,M_\omega^2,M^2\right)
   -6 M_\omega^2 M^2 B_0\left(M_\rho^2,M_\omega^2,M^2\right)
   \nonumber\\
   &+& 3 M_\rho^2 M^2
   B_0\left(M_\rho^2,M_\omega^2,M^2\right)+3
   M^4 B_0\left(M_\rho^2,M_\omega^2,M^2\right)-6 M_\rho^4
   B_0\left(M_\rho^2,M_\omega^2,M^2\right)\nonumber\\
&+& 3
   M_\omega^4 B_0\left(M_\rho^2,M_\omega^2,M^2\right)+A_0\left(M_\omega^2\right) \left(-3 M_\omega^2+6
   M_\rho^2+3 M^2\right)\nonumber\\
&+& 3 A_0\left(M^2\right) \left(M_\omega^2+2 M_\rho^2-M^2\right)-6 M_\omega^2 M_\rho^2+5 M_\rho^4-6
   M_\rho^2 M^2\Biggr\}\,.
\label{waveFROmegaloop}
\end{eqnarray}

\medskip

The contributions of the one-loop diagrams to $f_1(0)$ are:
\begin{eqnarray}
f_1[1] & = & \frac{e \left(-3 \left(M_\rho^2
+2 M^2\right) B_0\left(M_\rho^2,M^2,M^2\right)+6 A_0\left(M^2\right)+M_\rho^2-6
   M^2\right)}{288 \pi ^2 F^2} \,,\nonumber\\
f_1[2] & = & -\frac{e \left(M_\rho^2 \left(495
   B_0\left(M_\rho^2,M_\rho^2,M_\rho^2\right)+196\right)-
   534 A_0\left(M_\rho^2\right)\right)}{2304 \pi ^2 F^2} \,,\nonumber\\
f_1[3] & = & 0\,,\nonumber\\
f_1[4] & = & 0\,,\nonumber\\
f_1[5] & = & 0\,,\nonumber\\
f_1[6]+f_1[7] & = & 0\,,\nonumber\\
f_1[18]+f_1[9] & = & 0\,,\nonumber\\
f_1[10]+f_1[11] & = & -\frac{e \left(M_\rho^2 \left(10-99
   B_0\left(M_\rho^2,M_\rho^2,M_\rho^2\right)\right)+6
   A_0\left(M_\rho^2\right)\right)}{768 \pi ^2 F^2} \,,\nonumber\\
f_1[12] & = &  \frac{e A_0\left(M^2\right)}{16 \pi ^2 F^2}\,,\nonumber\\
%f_1[29]+f_1[30] & = & 0\,,\nonumber\\
f_1[13] & = & -\frac{e A_0\left(M^2\right)}{16 \pi ^2 F^2} ,\nonumber\\
f_1[14] & = & 0,\nonumber\\
f_1[15] & = &  \frac{e g_{\omega \rho \pi}^2}{288 \pi ^2 F^2 M_\rho^2} \Biggl\{3
   A_0\left(M_\omega^2\right) \left(-2 M_\omega^2+M_\rho^2+2
   M^2\right)\nonumber\\
&+& 3 A_0\left(M^2\right)
   \left(2 M_\omega^2+M_\rho^2-2 M^2\right)+M_\rho^2 \left(-3 M_\omega^2+4 M_\rho^2-3
   M^2\right)\nonumber\\
&-& 3 \left[M^2 \left(4
   M_\omega^2+M_\rho^2\right)+M_\omega^2
   M_\rho^2+M_\rho^4-2 M^4-2
   M_\omega^4\right]
   B_0\left(M_\rho^2,M_\omega^2,M^2\right)\Biggr\},\nonumber\\
f_1[16]+f_1[17] & = & 0,\nonumber\\
f_1[18]+f_1[19] & = & \frac{e g_{\omega \rho \pi}^2}{288 \pi ^2 F^2 M_\rho^2} \Biggl\{3
   A_0\left(M_\omega^2\right)
   \left(M_\omega^2+M_\rho^2-M^2\right) \nonumber\\
&+& 3
   A_0\left(M^2\right)
   \left(-M_\omega^2+M_\rho^2+M^2\right)+M_\rho^2 \left(-3 M_\omega^2+M_\rho^2-3
   M^2\right) \nonumber\\
&-& 3
   \left(M_\omega^4-2 M_\omega^2 \left(M_\rho^2
   +M^2\right)+\left(M_\rho^2-M^2\right)^2\right)
   B_0\left(M_\rho^2,M_\omega^2,M^2\right)\Biggr\} ,\nonumber\\
f_2[20] & = &  0\,.
\label{loopdiagramcontf1}
\end{eqnarray}

The contributions of the one-loop diagrams to $f_2(0)$ are:
%diagrams
%separately (contained in file ''RhoMMResultsByJambul'' in
%Directory ''RhoMagneticMoment'' compared with Dalibor, Correct)
\begin{eqnarray}
f_2[1] & = & -\frac{e}{288 F^2 \pi ^2}
\left\{M_\rho^2-6 M^2+6 A_0\left(M^2\right)+6
\left(M_\rho^2-M^2\right)
B_0\left(M_\rho^2,M^2,M^2\right)\right\}\,,\nonumber\\
f_2[2] & = & \frac{e}{2304 F^2 \pi ^2}
\left\{\left[136-513
B_0\left(M_\rho^2,M_\rho^2,M_\rho^2\right)\right] M_\rho^2+546
A_0\left(M_\rho^2\right)\right\}\,,\nonumber\\
f_2[3] & = & 0\,,\nonumber\\
f_2[4] & = & -\frac{3\,e}{64 F^2 \pi ^2}\,\left\{4
M_\rho^2-3
A_0\left(M_\rho^2\right)\right\}\,,\nonumber\\
f_2[5] & = & 0\,,\nonumber\\
f_2[6]+f_2[7] & = & 0\,,\nonumber\\
f_2[8]+f_2[9] & = & 0\,,\nonumber\\
f_2[10]+f_2[11] & = & \frac{e}{768 F^2 \pi ^2}\,
\left\{M_\rho^2 \left[99
B_0\left(M_\rho^2,M_\rho^2,M_\rho^2\right) -10\right]-6
A_0\left(M_\rho^2\right)\right\}\,,\nonumber\\
f_2[12] & = & \frac{e}{8 F^2 \pi ^2}\, A_0\left(M^2\right)\,,\nonumber\\
%f_2[29]+f_2[30] & = & 0\,,\nonumber\\
f_2[13] & = & -\frac{e}{8 F^2 \pi ^2}\,
A_0\left(M^2\right),\nonumber\\
f_2[14] & = & 0,\nonumber\\
f_2[15] & = &  -\frac{e g_{\omega \rho \pi}^2}{576
   \pi ^2 F^2 M_\rho^2} \Biggl\{-6
   A_0\left(M_\omega^2\right)
   \left(M_\omega^2+M_\rho^2-M^2\right)\nonumber\\
   &+& 6 A_0\left(M^2\right) \left(M_\omega^2+2
   M_\rho^2-M^2\right)+M_\rho^2 \left(-3
   M_\omega^2+M_\rho^2-3 M^2\right)\nonumber\\
&+& 6 \left(M_\omega^4-2 M^2
   \left(M_\omega^2+M_\rho^2\right)+M_\omega^2
   M_\rho^2+M_\rho^4+M^4\right)
   B_0\left(M_\rho^2,M_\omega^2,M^2\right)\Biggr\},\nonumber\\
f_2[16]+f_2[17] & = & 0,\nonumber\\
f_2[18]+f_2[19] & = & \frac{e g_{\omega \rho \pi}^2}{288 \pi ^2 F^2 M_\rho^2} \Biggl\{3
   A_0\left(M_\omega^2\right)
   \left(M_\omega^2+M_\rho^2-M^2\right) \nonumber\\
&+& 3
   A_0\left(M^2\right)
   \left(-M_\omega^2+M_\rho^2+M^2\right)+M_\rho^2 \left(-3 M_\omega^2+M_\rho^2-3
   M^2\right)\nonumber\\
&-& 3
   \left(M_\omega^4-2 M_\omega^2 \left(M_\rho^2+M^2\right)+\left(M_\rho^2-M^2\right)^2\right)
   B_0\left(M_\rho^2,M_\omega^2,M^2\right)\Biggr\},
\nonumber\\
f_2[20] & = & -\frac{e\,g_{\omega \rho \pi}^2}{64 \pi ^2 F^2
\left(M_\omega^2-M ^2\right) } \left[M_\omega^4-2
A_0\left(M_\omega^2\right) M_\omega^2-M ^4+2 M ^2 A_0\left(M
^2\right)\right]
 \,.
\label{loopdiagramcontf2}
\end{eqnarray}
\end{appendix}

%\newpage

\end{document}